\documentclass[usenatbib]{basi}
\usepackage[T1]{fontenc}
\usepackage[british]{babel}
\usepackage[varg]{txfonts}
\usepackage{rotating}
\usepackage{dcolumn}
\begin{document}

\title[Spectroscopy of novae]{Spectroscopy of Novae -- A User's Manual}
\author[S. N. Shore]%
       {Steven N. Shore$^{1,2}$\thanks{email: \texttt{shore@df.unipi.it}}\\
      $^1$Dipartimento di Fisica ``Enrico Fermi'', Universit\'a di Pisa, largo B. Pontecorvo 3, Pisa 56127, Italy\\
       $^2$INFN - Sezione di Pisa}

\pubyear{2012}
\volume{00}
\pagerange{\pageref{firstpage}--\pageref{lastpage}}

\date{Received --- ; accepted ---}

\maketitle
\label{firstpage}

\begin{abstract}
The spectroscopic development of classical novae is described as a narrative of the various stages of the outburst.  The review highlights the multiwavelength aspects of the phenomenology and the recent developments related to structure, inhomogeneity, and dynamics of the ejecta.   Special emphasis is placed on the distinct behavior of the symbiotic-like recurrent novae.  
\end{abstract}

\begin{keywords}
 (stars:) novae, cataclysmic variables;  stars: winds, outflows;  radiative transfer; hydrodynamics; (stars:) supernovae: general 
\end{keywords}

\section{Introduction}\label{s:intro}

At the outset of the twentieth century, Agnes Clerke's {\it Problems in Astrophysics}  (Clerke 1902) illustrated the promise and progress of the then nascent discipline by emphasizing the centrality of spectroscopy.  This was typical of the ``astrophysics'' movement, led by such notables as Hale, Tacchini, Hale, Secchi, Voegel, and Huggins. They promoted the growth of astronomical investigations based on spectroscopic analyses by founding journals, promoting conferences, and national and international societies.   Among her choices, she described there and elsewhere the bright Galactic nova T Aur 1892 .  This bright Galactic nova was discovered in outburst with a spectrum that resembled a giant (although the classification was not yet in vogue), dominated by absorption lines and Balmer emission.  It then entered a period of solar constraint when it was not observable and, on emergence, Clerke recounts that it spectrum had completely changed.  It resemble a nebula but the widths of the lines, that appeared more like bands, were like those of a Wolf-Rayet star without the absorption troughs.  The change was repeated in a number of others.   One phrase from chapter on ``temporary stars''  the stands out as almost a clarion call for future research: `` Nova Auriga was neither the first, nor the last, temporary star to don, in fading, the garb of nebular light.''  The same was noted for Q Cyg and GK Per.   In mid-century, Cecilia Payne-Gaposchkin produced her {\it vade mecum} {\it The Galactic Novae} (1957) that summarized the disparate studies of individual objects, largely conducted at diverse observatories and much of it published only in bits and pieces.   More important, perhaps, is her chapter on ``the geometry of the nova spectrum''.   The aim of this review is to explain, in general in qualitative terms, the physical processes behind the observations that update these earlier works.  In specific illustrative cases, I will give more quantitative details.  But since novae have an inherently diverse bestiary, each passes through each of these stages at a different time and with different relative duration.  Yet it is significant that novae are, at their core, comparatively simple although seeing the forest through the trees has taken well more than a century.

The aim of this review is to provide an outline of some of the physical insights provided by reading the line profiles in novae.  I hope the reader will forgive a pedagogical approach but the literature is vast and examples are both numerous and widely dispersed.  The best first step is still the collection edited by Bode \& Evans (2008), the homage to the tradition founded by Payne-Gaposchkin's monograph.  If this encourages further personal exploration then the review will have fulfilled its goals.

\section{Classical novae as explosions: a theoretical contentexualization}

Classical novae are explosive mass ejection events caused by the ignition of a nuclear reaction on the surface on an accreting white dwarf (WD) under degenerate conditions.  The main nuclear reactions, CNO-hydrogen burning, goes ``supercitical'', in the form of a thermonuclear runaway (TNR), because the environment cannot initially respond dynamically to the rapid rise in the energy generation rate and the consequent heating.   The TNR, proceeding far from equilibrium, overproduces  critical, short-lived $\beta$-unstable isotopes, the most important of which is $^{15}$O.  These decay in about 100 sec and heat the accreted gas far more effectively that the radiative and conductive processes.  When the temperature finally rises above the degeneracy line, the layer above the nuclear zone is expelled.   The WDs come in two ``flavors'', ONe (from massive progenitors) and CO (from slightly lower mass progenitors) and these have been distinguished in the abundance patterns derived from the spectroscopy at late times when the expanding shell is optically thin.   The maximum luminosity and nucleosynthesis depend the WD  mass and composition.  In this standard model, the more massive WD requires a lower accumulated mass to reach the critical envelope pressure at which the TNR is ignited and, consequently, the highest luminosities, lowest ejecta masses, and highest velocities should be from degenerates near the Chandrasekhar limit; these have also been tapped as possible SN Ia precursors depending on their parent cataclysmic binary system and are the recurrent novae that have inter-explosion intervals of decades instead of millennia (e.g. Bode 2010).   One dimensional models of the accretion and ignition show that the mass ejected can be as large as $10^{-5}$M$_\odot$ with velocities of order a few thousand km s$^{-1}$.  The essential picture of the initiation is, consequently, well established (Starrfield et al. 2008; Starrfield et al., this issue).  In addition, since this is not destructive of the white dwarf, the explosions will repeat once sufficient mass again accumulates.  Thus, I will not distinguish between the recurrent systems and those that have been seen in only one outburst.   New multidimensional models provide the explanation for the deep mixing, that on the start of convection in the thin ignition zone deep penetrating vortices are formed by a buoyant instability in plumes at the core-envelope interface that effectively dredge up the underlying matter (e.g. Casanova et al. 2011) and produce a dispersion of abundances within the ejecta as a natural consequence of the dynamical convection. 

\section{The spectroscopic phases and their diagnosis}

An important point to keep in mind is that nova ejecta are, essentially, passive media.  This is not true, as I'll explain later, for a specific subgroup -- the symbiotic like systems, for which the ejecta and their associated shock propagates through a dense circumstellar medium -- but otherwise the ejecta are a sort of complex screen through which light from the central white dwarf passes.  In many ways, they resemble the classical nebular radiative transfer problem since the requirements of thermal and mechanical equilibrium are both relaxed. But an essential difference, the strong time dependence of the optical depth as a function of energy, makes the spectral formation more complicated.

In the interstellar medium, the notion of a Str\"omgren sphere serves as a point of departure for photoionization analyses.  A central source illuminates a diffuse, rarified circumstellar medium.  Depending on the spectral energy distribution of the central source, the surroundings are ionized or merely excited by resonance lines.  In a gas of sufficiently low density, specific to each transition and depending on the collision rates (hence electron temperature and density) and the spontaneous transition probability of the specific transition, the photon is either scattered resonantly or absorbed or emitted.  In general, the temperatures in nova ejecta are too high for molecules to be determinative in the energy balance although they are detected in the earliest spectra (see below).  Instead, to concentrate on the most familiar example, the hydrogen Lyman and Balmer lines, the emission is either from UV absorption in the Lyman series and emission through radiative de-excitations in the Balmer (and higher) series or from recombination.  Collisional excitation is generally negligible.   In a static H II region, the statistical balance achieved in relatively short time ionizes the surroundings to a distance that depends uniquely on the density (through the recombination rate) and the incident flux in the appropriate ionizing continuum.  The expansion that results from the overpressure of the region is sonic and the ionized region grows by engulfment while maintaining at any moment statistical equilibrium.

In a stellar wind, a similar situation holds when the radial outflow maintains a constant velocity profile.  Advection of any parcel of gas can be computed hydrodynamically and, since the flow is supersonic, the populations are determined by local processes that are not regulated by hydrostatic equilibrium.  We will go deeper into this later in this discussion.

Novae are more complicated for one reason: the ejecta are in free expansion and the density (and therefore ionization  and temperature) is systematically changing in time.   And because of this, although this review will not treat the details of the light curves, this phenomenology is intimately connected with the evolution of the spectrum of the ejecta.

\subsection{Phase 1: the fireball}

The first stage of the expansion, the {\it fireball}, has its closest analogy with a standard terrestrial nuclear explosion.\footnote{Outstanding examples of nuclear test films, showing the first few seconds of the fireball expansion, are now available through {\it YouTube}.  The reader is encouraged to watch these, they are singularly instructive, if not also disturbing.}  The medium is heated by the passage of the shock and at its highest temperature while in free expansion and rapidly cooling.  Depending on the initial temperature, the expansion time is much shorter than the radiative cooling so the ejecta are essentially adiabatic.  Whatever the initial temperature, the rate of expansion depends only on the maximum velocity and the optical depth is from bremsstrahlung and electron scattering.  This phase has been directly observed in very few novae, it requires an alert while the nova is still rising to maximum light. Consequently, its connection with the optical light curve is poorly understood.  Some novae show a pre-maximum halt before the final rapid rise to maximum.  It is plausible that this is the point at which the fireball cools to the critical temperature at which a recombination wave begins, setting the stage for the next spectral transition.  This stage in V1974 Cyg is shown the first two panels of Fig. 1 and the first panel in Fig. 2.  One important note: the first spectra cover a period of only about three days.   The photometric signature, shown in Fig. 2, is the almost discontinuous drop in the UV that coincides with the rise to maximum visible light.

The radio spectrum, mm and cm wavelengths, displays the simplest phenomenology but also illustrates the difference between continuum and line transfer, a central point in our discussion.  Because the predominant opacity at these wavelengths is continuous, i.e. thermal bremsstrahlung, the radiative transfer is especially simple.  The evolution of the monochromatic intensity is independent of the velocity gradient. Instead, the  rate is determined by the maximum expansion velocity and the optical depth by the density gradient.  The emission is characterized by an inverse power law and departure from a blackbody spectrum.  Its time development is, however, dependent on the recession of the photosphere within the ejecta as the density drops.  If the medium is isothermal, the change of the radial optical depth offsets the continued expansion of the outer envelope.  The frequency dependence of the opacity, $\kappa_\nu \sim \nu^{-3}$ means that at each frequency the time of  peak emission differs with the lower frequencies becoming transparent at later times.  This leads to a characteristic variation that depends only on the mass of the ejecta and the maximum expansion velocity.  One thing must, however, be kept in mind regarding the determination of distances and rates of expansion from multiwavelength light curves and imaging: the specific wavelength regime is sensitive to different portions of the ejecta and the appropriate mean expansion velocity must be used in the interpretation.  Fo instance, the optical lines, P Cyg or otherwise, are always lower velocity than those obtained from the resonance lines in the UV and weighted toward the denser, slower moving parts of the ejecta.  The high opacity of the ejecta in the radio, especially during the fireball stage, will weight that part toward higher velocities.  Using one regime to interpret another is ill-advised and can lead to systematic, although not obvious, errors.  Once the ejecta are resolved, however, the emission line velocities appropriate for the specific wavelength at which the images are obtained will permit accurate assessments.

\begin{figure}
\centerline{\includegraphics[width=14cm]{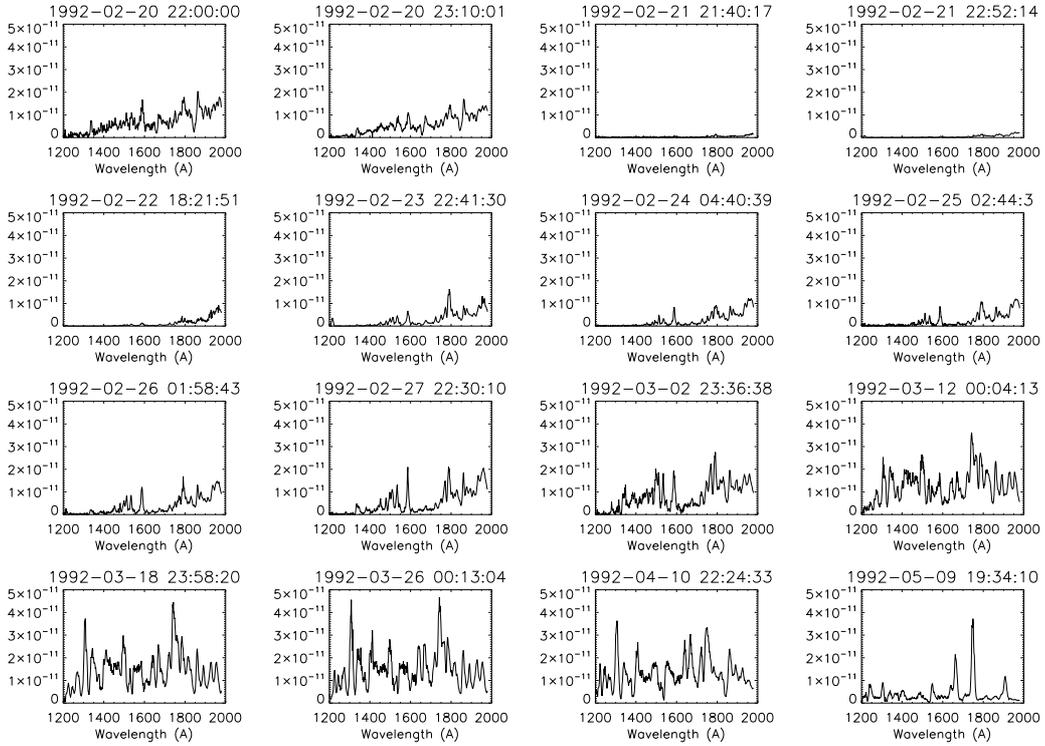}}
\caption{V1974 Cyg, 1200-2000\AA\ during the first months of the outburst.  This sequence, a unique record of the event, shows the fireball spectrum in the first day (one day after optical discovery) and the dramatic effect of the ``Fe curtain'' in less than one day.\label{f:one}}
\end{figure}

\begin{figure}
\centerline{\includegraphics[width=10cm,angle=90]{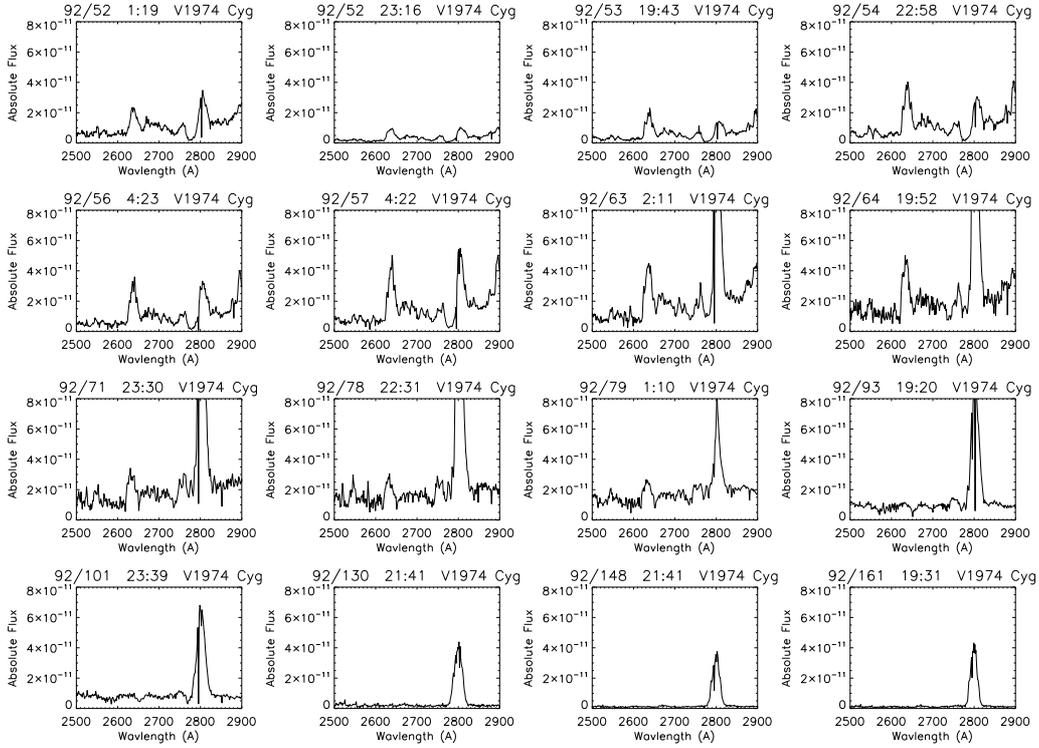}}
\caption{The 2000-3000\AA\ high resolutiuon variations of V1974 Cyg.  The Mg II 2800\AA\ line shows the transition from a low ionization overlying Fe-peak absorption curtain to the weak continuum and the domination of the spectrum by the ejecta. \label{f:one}}
\end{figure}


\subsection{Phase 2: the ``iron curtain''}

During the cooling stage, when the optical spectrum is strengthening, the ultraviolet -- the site of the resonance transitions of the same iron peak ions that dominate the optical spectrum -- suddenly becomes optically thick.  This is, for all practical purposes, a phase transition.  This is seen immediately after the fireball spectra in Figs. 1 and 2.  The medium has recombined at this stage and the main opacity source, the metal line complexes in the Balmer continuum, are now opaque. These connect to hundreds of  higher levels, all of which can radiatively de-excite through literally millions of lines.  These are all cross coupled, since the random walk of a photon through one line profile can couple it to any overlapping transition within the ejecta because of the velocity gradient.  The same is true for a stellar wind but here the medium is time dependent so the continuum optical depth of the entire volume is systematically changing throughout the time interval.  To complicate matters there are also collisional coupling between these levels so the treatment requires a fully NLTE analysis to completely model the spectra.  While the majority of the  transitions can be handled approximately in LTE, tens of thousands of strong lines require exact treatment (an early, but detailed physical study is found in Hauschildt et al. 1997).

\begin{figure}
\centerline{\includegraphics[width=11cm]{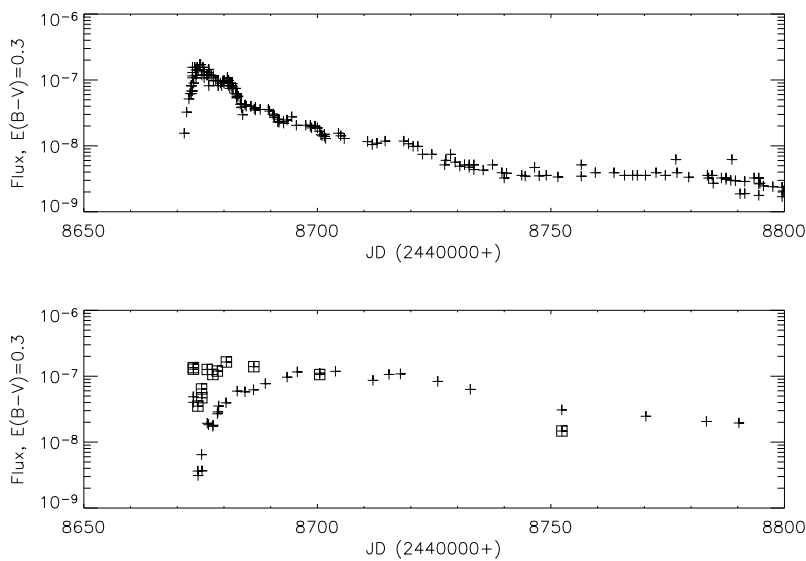}}
\caption{The photometric variations of V1974 Cyg during outburst.  Top: V magnitude.  Bottom: 1200-2000\AA\ (cross), 2000-3000\AA\ (square).  The rapid drop in the UV flux at the start is the initiation of the Fe-curtain stage, shown in Fig. 1, the point at which the UV becomes optically thin roughly corresponds to t$_3$ in the optical light curve.  The flux was bolometrically constant for the initial three months.\label{f:one}}
\end{figure}

Observations of T Pyx 2011 appear to hold the key to the phenomenon (e.g. Shore et al. 2011b).  In this nova, when first observed on the rise to maximum light, the optical displayed He I  P Cyg lines along with the Balmer series and little else.  As the optical continued rising, the He I disappeared and was replaced by a forest of complex, broad iron peak lines whose P Cyg profiles had absorption at lower velocities (Fig.4).  These absorption components, beginning at around -800 km $^{-1}$, showed a systematic displacement toward larger blueshift and a strengthening of the emission on the low ionization states.  This behavior has been reported in other novae, mainly before the era of CCD spectroscopy (so it was suspected to be, perhaps, an artifact of the measurement techniques) but spectrophotometric measurements verify the reality of the phenomenon.    The lines in the optical were marginally transparent while those in the UV were strongly absorbed, the Fe-curtain completely dominated that part of the spectrum.  Thus, initially, the strong pumping from the UV produced comparatively broad absorption lines.  But as the ejecta thinned out, and the UV became more optically thin, individual transitions coupled less and the envelope broad line disintegrated into a forest of individual, weaker features that gradually faded.  To understand this, consider the coupling between two lines that are separated by some velocity $\Delta v$.  This corresponds to different radial distances in the ejecta so the two can be coincident and strongly absorb together, hence mutually shadowing each other.  But if they are well separated, the continuum can be seen by both and the levels may be more strongly pumped.  This depends on their relative $gf$ values and excitation energies.   The huge number of possible transitions insures that coincidences will be encountered throughout the optically thick stage in the UV and, therefore, that the relative visibility of the absorption lines will change over time.  The presence of these individual narrow components demonstrates the early formation of fine structure in the ejecta, a problem hat remains unsolved in novae.  Unlike O stars, for which campaigns such as MUSICOS showed sector structure connected to rotation (and possibly magnetic fields), it is unlikely that either plays a role in the formation and variation of these narow lines.

\begin{figure}
\centerline{\includegraphics[width=9.5cm,angle=90]{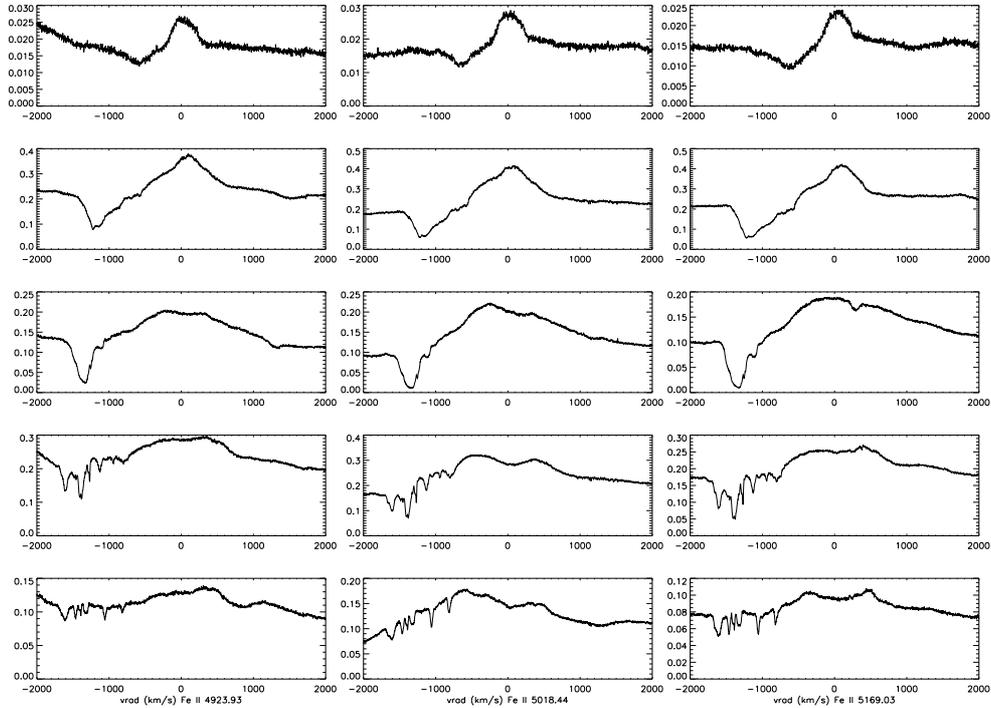}}
\caption{Fe II RMT 42 during the 2011 outburst of T Pyx, showing the evolution of the narrow components as the ejecta first recombine and then thins on expansion.  Three different lines of the multiplet are shown in the columns, time increases from top to bottom (see Shore et al. 2011b).\label{f:one}}
\end{figure}

\begin{figure}
\centerline{\includegraphics[width=10cm,angle=90]{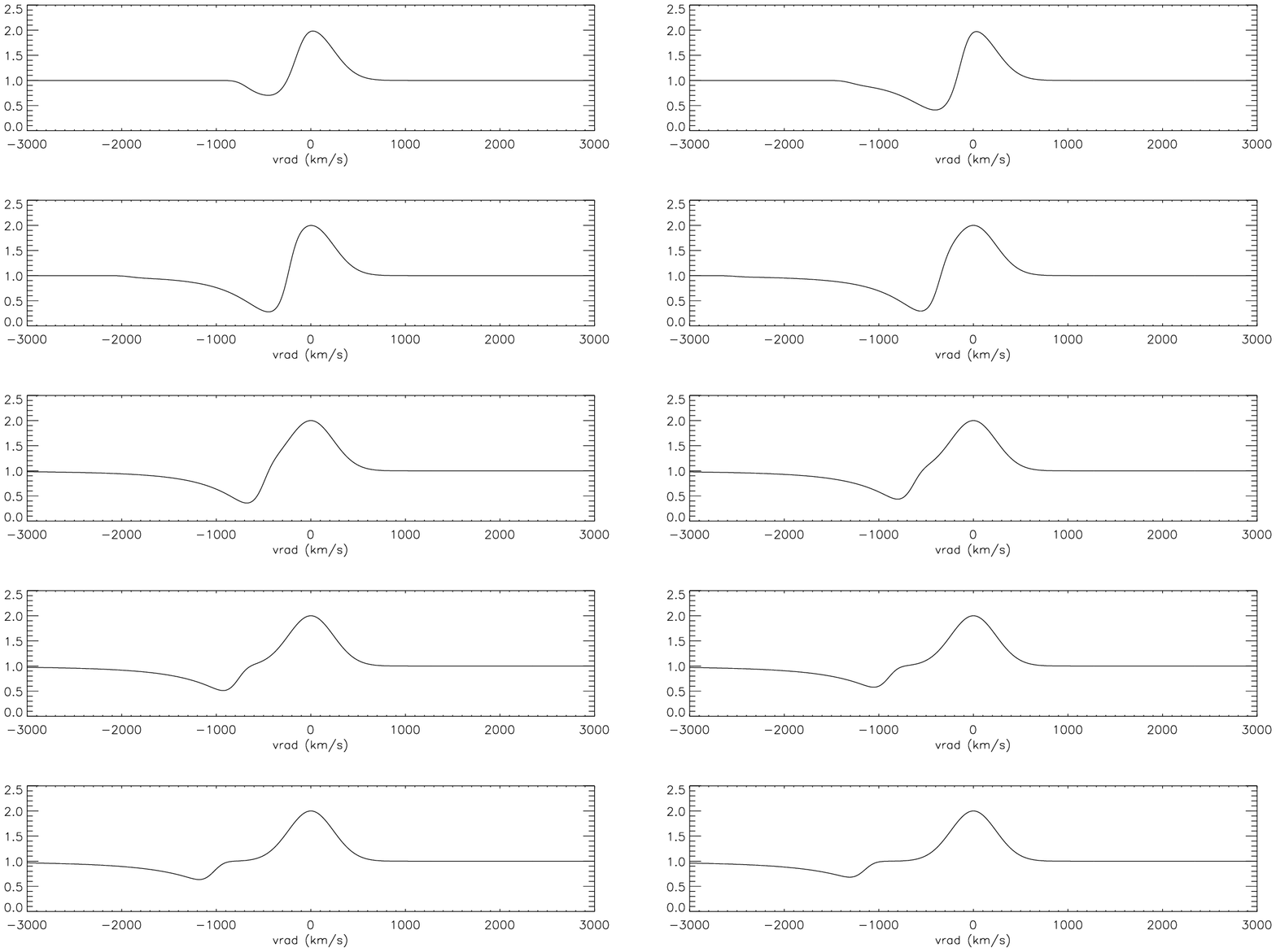}}
\caption{A simple model showing the propagation of a recombination wave through the ejecta at the start of the Fe-curtain phase.  Each panel is a time step scaled to the recombination time, in this case about one day for each panel.  The time increases from right to left and from top to bottom (choose how you want to pass through the figure).  The ejecta are assumed to have a balliistic radial velocity and density dependence. \label{f:one}}
\end{figure}

Much of this should be quite familiar to the reader.  In H II regions, for instance, the two canonical cases of recombination, so called case A and case B, differ in the optical depth of the Lyman transitions.  Those in case B are optically thick, hence a change in the relative line intensities because of the available radiative excitation channels in the UV and line trapping.  A velocity gradient changes this.  The hydrogen lines, being sparsely distributed through the spectrum, become more transparent in their cores because of differential shifts.  The photons emerging in the core at one depth see a reduced opacity above them due to the Doppler shift and this changes the escape probability.  But if the medium has a large stochastic velocity field, or a small gradient (defined relative to the random component) then the line ratios follow case B.  Novae are, in this way, very high velocity gradient H II regions or, even better, planetary nebulae at high speed.   The ejecta are a filter, a passive medium whose ionization and opacity has no effect on the dynamics since the gas is already in supersonic expansion.  This is not true for a wind since the velocity gradient asymptotically levels out.

On this last point, there remains a debate on the presence of a wind during or following the  explosion.  In some models, there is no true explosion, just a TNR that provokes a super-Eddington phase that drives a strong, continuous outflow.  This is so optically thick that it appears as an expanding photosphere and this is said to account for the full phenomenology.  There are several important points here that should be clarified.  {\it Any expanding medium has a pseudo-photosphere while opaque}.  Only a nebular spectrum, and then only in certain wavelength intervals, will be truly transparent.   There is, in this sense, no difference between a wind and freely expanding ejecta.  The difference is in the details.  A wind starts at rest, at a stationary surface that is connected to the rest of the central body by the requirement of hydrostatic equilibrium.  In other words, the material that is ultimately accelerated to escape velocities must start from rest with no inner boundary.  This also means that the flow has an asymptotically constant velocity at large distance, we'll call this ``infinity'' for convenience.  The flow is either mechanically driven by some stress, be it pressure fluctuations or nonlinear MHD wave coupling.  Other possibilities come to mind for stars but these are the two extremes.  For the weak wind case, internal heating by a wide variety of processes can drive the flow above the thermal speed (i.e. sound speed) and produce a continuous outflow.  Thus originates the solar wind, for instance.  But it has long been understood that radiation pressure, especially when coupled to the medium through line transitions, is remarkably effective as an accelerator.  Thus, the spectrum is intimately connected with the dynamics and the velocity gradient -- the controlling parameter in the opacity once the velocity is super-thermal -- determines the radiative acceleration that, {\it mutatis mutandis}, changes the velocity gradient.  The line profiles resulting from an isothermal outflow are iconographic, showing well defined blueshifted edges to the absorption.  This is the signature of the approach to asymptotic expansion velocity, the opacity along any line of sight is large because the velocity gradient becomes vanishingly small and the radiative transfer is strongly non-local.  But this does not affect the velocity gradient since, locally, this is at the outer boundary of the flow where the density is low so the escape probability along anything but the radial direction is greatly enhanced.

Now compare this with the expectations for a ballistic or explosive ejection.  The velocity never becomes constant, the whole point of ballistic flow is that the instant some matter is ejected it continues with its initial velocity.  Thus, a higher initial speed means the material is farther at any moment and the velocity gradient is statistically constant.  There is no outer enhancement of the optical dept.  On the contrary, the optical depth systematically decreases with distance.  Consequently, there is no ``edge'' to the absorption trough, even with the expansion of the recombination front (Fig. 5).  The profile extends smoothly to the continuum level at maximum expansion velocity.  Individual narrow absorption features, resulting from UV excitation within the curtain, change opacity systematically but there  is no accelerating advection within the ejecta.  A piece of the matter remains at constant radial velocity, it's the local opacity that changes and the accelerations are only apparent.

\subsection{Phase 3: the lifting of the ``iron curtain''}

The next stage, connected with a distinct moment in the optical photometry, is when the UV turns optically thin in the lines.  Two things happen simultaneously.  The loss of absorption in the UV means the outer ejecta are exposed to a progressively harder radiation field (Figs. 6, 7).  This is not only from the lines, however, the ionization increases because the FUV continuum absorption decreases with decreasing density.  This drives an ionization front outward through the ejecta as the photosphere color temperature rises, ionizing the ejecta and removing the species responsible for the curtain.  But the recombination rate also drops, hence the ionization falls out of balance with recombination.  This further enhances the ionization and, because of the feedback, the transition progresses rapidly.  Slow ejecta take some time for this, also massive ejecta are systematically more opaque for longer.  But it is an inevitable consequence of the ever-present central white dwarf, the ejecta must not only dramatically ionize -- up to the effective temperature of the central source -- but also turn transparent.  The flux peak then shifts to the UV and FUV, and the optical fades systematically.

\begin{figure}
\centerline{\includegraphics[width=12cm]{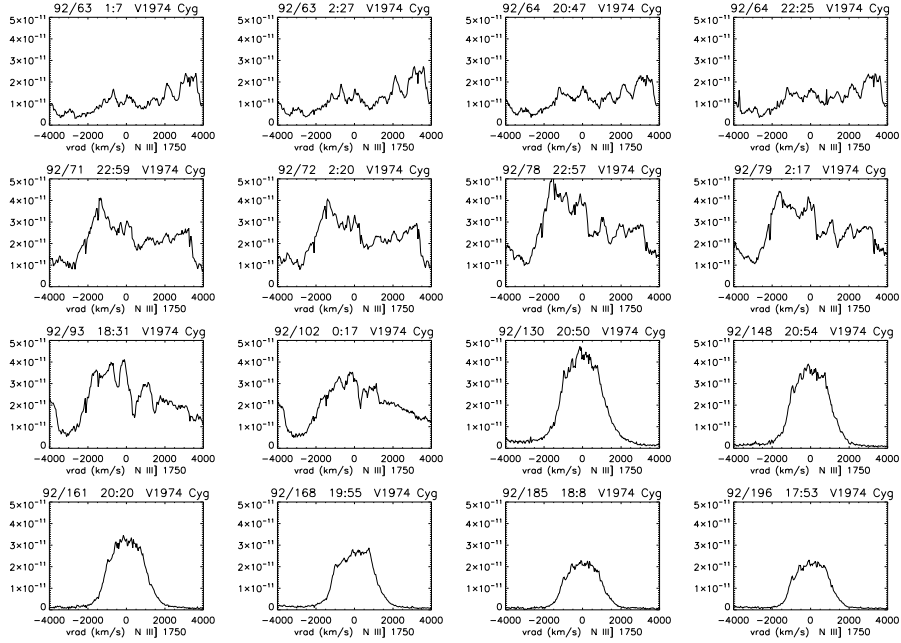}}
\caption{Variations in the resonance N III] 1750\AA\  line for V1974 Cyg during the first year after outburst.  Note here and for Mg II how the emission line emerges as the Fe-curtain lists (IUE spectra, high resolution).\label{f:one}}
\end{figure}

\begin{figure}
\centerline{\includegraphics[width=12cm]{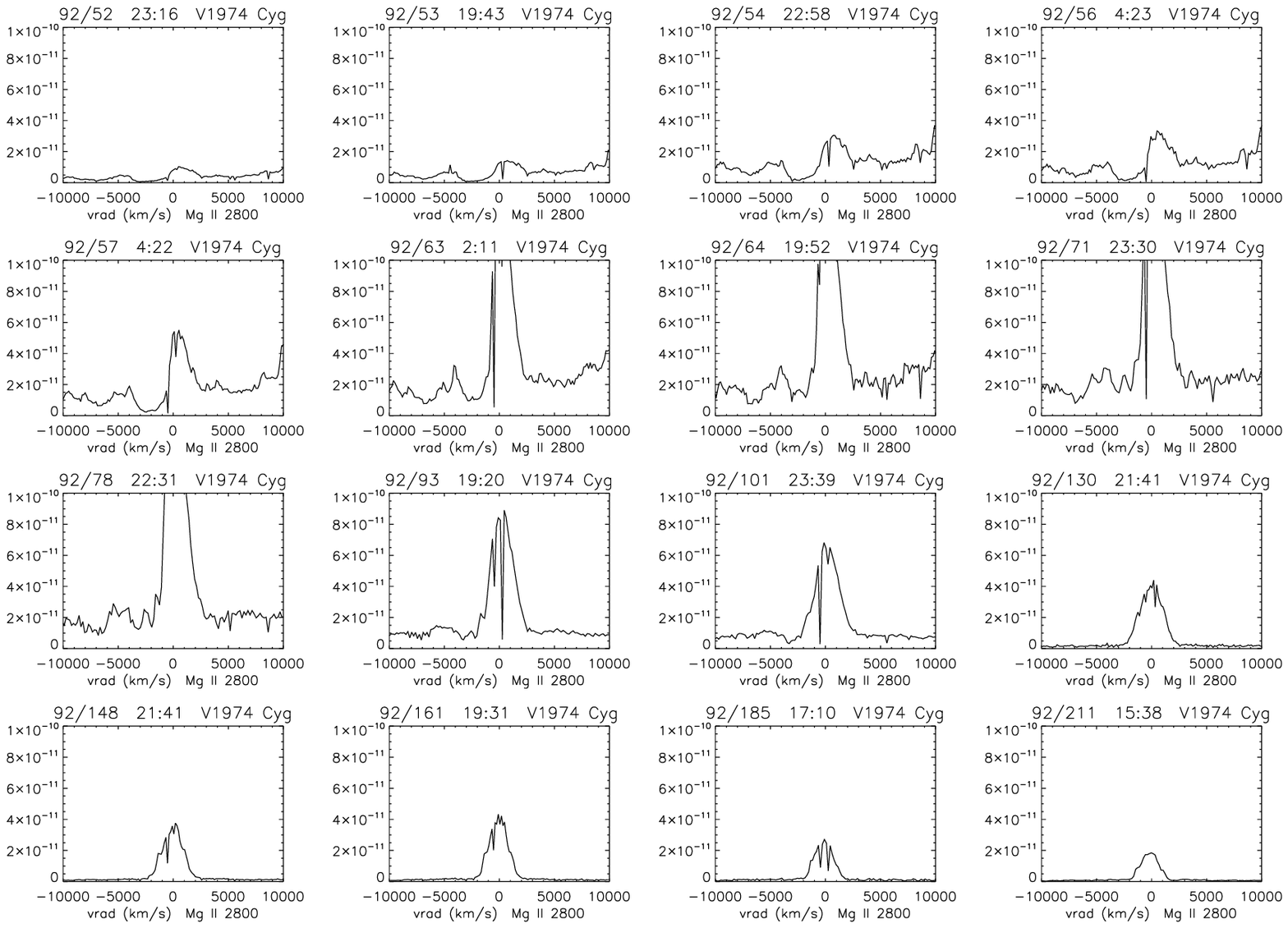}}
\caption{The Mg II 2800 line evolution for V1974 Cyg during outburst (IUE spectra, high resolution).\label{f:one}}
\end{figure}

\subsection{Phase 4: the transition stage}

Once the UV turns transparent, the tell tale nebular lines so familiar in the late stages of the expansion begin to appear (e.g. 
Moro-Martin et al. 2001, Della Valle et al. 2002).  The first signal is enhanced He II emission relative to the N III 4640\AA\ complex.  Accompanying the transition is the first appearance of the [O III] 4363, 4959, 5007 \AA\ lines.  These are accompanied by their isoelectronic counterparts [N II] 5755, 6548, 6583 \AA, although the last two are usually blended with H$\alpha$ and only visible later.  Since these are analogs, the [O III] and [N II] profiles can be used to determine the spatial (velocity) dependent electron density and temperature uniquely if  both ions are visible.  This was done recently for the late spectrum T Pyx (about one year after outburst) with the somewhat surprising result that the electron temperature was about 40 kK with n$_e \approx 10^6$ cm$^{-3}$ (a surprisinly high value that may indicate the presence of additioonal heating, e.g. internal shocks).  An advantage presented by the ejecta spectrum has not yet been sufficiently exploited: the large velocity gradient and simple ballistic radial dependence maps the location to single points in the line profiles.  In the nebular phase, and also during the transition, the optical depth of the ejecta is small and, therefore, the matching of structures in the line profiles indicates variations in the mass at a given velocity.  A ratio of the profiles for the various plasma diagnostic transitions then gives a spatially dependent density.  The Balmer line, H$\beta$, gives the filling factor so together it is possible to obtain a mass and filling factor.   The technique has been employed for interstellar lines from different ions, which requires typically much higher resolution (better than a few km s$^{-1}$) but the narrowest features in the ejecta emission profiles are usually 50 to 100 km s$^{-1}$ wide.  This is also true for the one spatially resolved UV/optical study, V1974 Cyg.  This last study also highlights one of the potential pitfalls in using integrated line fluxes to obtain abundances and plasma parameters.  There are hints of abundance variations in individual knots, something predicted by models of the outburst.  In the absence of spatially resolved ejecta, the knots in the profiles give an indication of this inhomogeneity and the individual structures can be followed through the expansion since they should remain stationary in velocity (again the point that the structures don't advect differentially as they would in a wind).

\begin{figure}
\centerline{\includegraphics[width=10cm,angle=90]{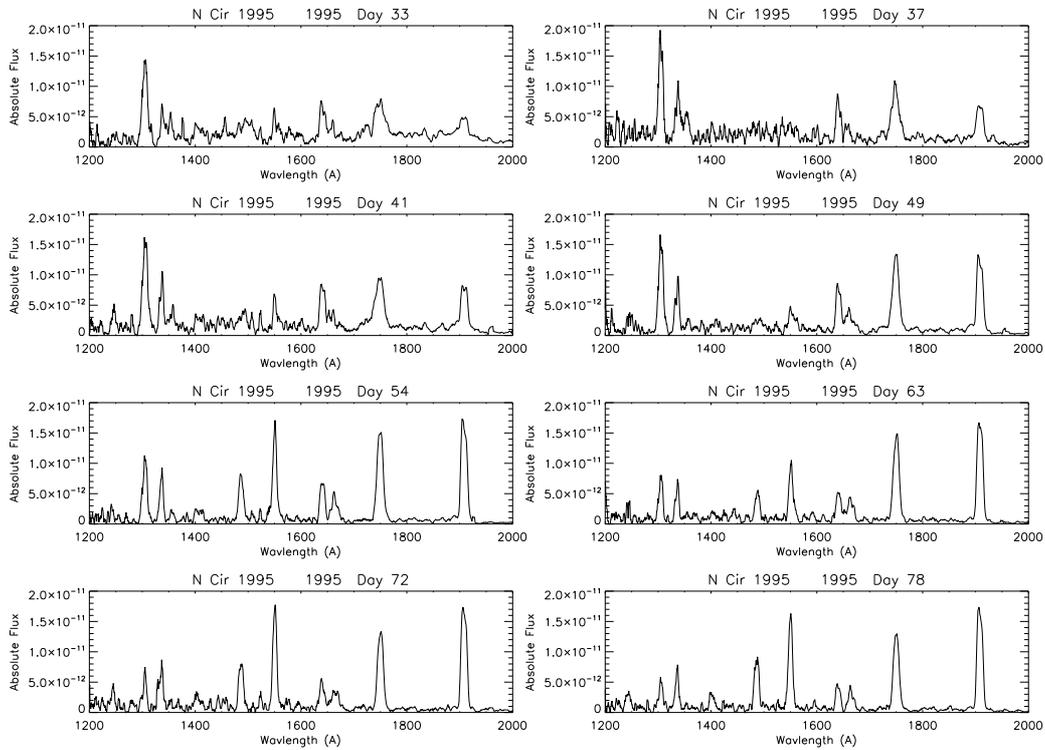}}
\caption{The transition stage of BY Cir 1995 showing the first appearance in the UV of low ionization emission lines.\label{f:one}}
\end{figure}

The optical line profiles also change, since they are no longer strongly coupled to those levels that were previously pumped by absorption in the UV resonance bands.  The lower density region, above the pseudo-photosphere, emit transparently and over a large solid angle. Thus, individual filaments and knots, prevously invisible, now appear on the profiles.  An example is shown in Fig. 8 for two ONe novae, LMC 2000 and V382 Vel, who will appear again later in our discussion.

\begin{figure}
\centerline{\includegraphics[width=10cm]{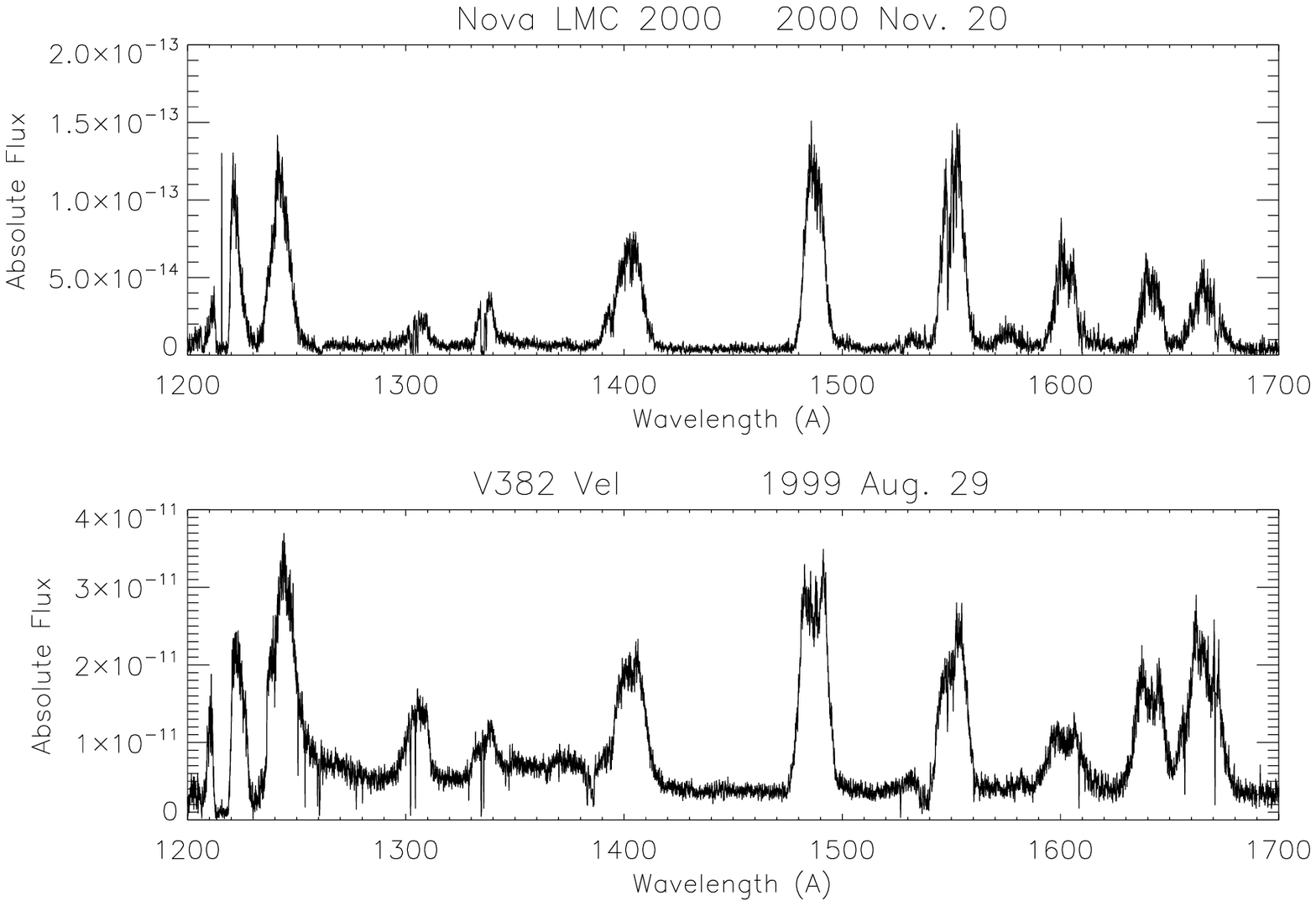}}
\caption{Comparison of two ONe novae, V382 Vel (Galactic) and LMC 2000 during the transition stage in the ultraviolet (HST/STIS spectrscopy).\label{f:one}}
\end{figure}

\subsection{Phase 4a. dust formation}

Although this is mainly distinguished by changes in the optical and infrared photometry, one distinctive feature is linked to the spectroscopy.  The rapid drop in the optical flux with a simultaneous increase in the mid-IR emission (longward of 2$\mu$ is the dead giveaway of dust condensation.  How the dust forms is a separate topic (see, e.g. Evans et al. 1997).  Empirically, it typically happens during the transition stage, about 100 days after optical peak (e.g. DQ Her).   In SN II this is accompanied by a weakening or disappearance of the redshifted portion of the ejecta lines.  This has not been seen optically in most classical novae but a related event gives important information on the site of the phase transition.  The CO nova V705 Cas was, by chance, observed in the UV with {\it IUE} during its dust forming event in late 1993.  The spectrum in the 1200-3000\AA\ region remained unchanged in line ratios and components but decreased by more than an order of magnitude in flux.  This was consistent with the flux increase in the IR and showed that the dust was forming in the outer portions of the ejecta (large distance) and was redistributing the absorbed flux (Shore et al. 1994) that was modeled with large grains since the additional extinction was grey.   Optical spectropolarimetry is the ideal addition to the analytic toolbox for separating the different contributors and detailing the structures of the dust forming regions but has been little exploited to date.  A lovely example of what can be done is shown by Kawabata et al. (2001) for V1494 Aql.   For other issues related to dust and chemistry, it is best to consult the review in this issue by Evans \& Gehrz.

\subsection{Phase 5: the nebular and coronal spectra}

The appearance of high ionization species, e.g. [Fe VI], [Fe VII], [Ca V], signals the coronal phase, so called because of the presence of lines familiar from the solar corona.  The  term is, however, misleading since the conditions are significantly different.  The corona is a magnetostatic atmosphere, maintained in loops and in the base of the solar wind by nonthermal heating  mechanisms.  The losses are radiative but the heating, both {\it in situ} and nonlocal, is from magnetic dissipation and shocks.  In nova ejecta, the main input is the X-ray and FUV emission from the WD in a dynamical medium.  Any photons intercepted that can ionize the gas will be weakly balance by recombination that controls the recombinations in a continually rarifying environment.  There are two phases that are not well distinguished at present.  In XRs, the initial emission is always hard, a source extending up to and beyond 10 keV.  This has a finite duration, perhaps several weeks.  The so-called {\it supersoft} component (see Ness, this issue), thought to be due to continuing nuclear processing in the WD envelope from the hydrogen that is not expelled in the explosion, is strongly modulated by the change in the column density of the ejecta with time.  This was well sampled for V1974 Cyg and now for other novae, mainly due to {\it Swift}.  The ionization of the ejecta increases and the opacity decreases, a reprise of the phenomenology of the Fe curtain, and in the space of several months the source dominates the state of the expanding gas.  The ionization eventually freezes-out when the rate of recombination is solely determined by the decreasing electron density.

\section{Geometry of the ejecta}

The interpretation of the structure of the unresolved ejecta has a long history, going back farther than the extended, prescient discussion by  Payne-Gaposchkin (1957).  It was realized quite early in the game that the stationarity of the individual emission peaks in the optically thin line profiles were not mere density differences but actual signatures of the geometry of the ejecta.  First interpreted as rings and cones, an approach made more precise by e.g. Hutchings (1972) and Gill \& O'Brien (1999)  to cite a few examples, can be generalized to indicate that axisymmetry is fundamental to the observed structure.  I will use the most recent observations and modeling of T Pyx 2011.  The examples show that bipolar cones with variable opening angles easily produce the same structures.  Depending on the inclination angle of the symmetry axis to the line of sight and the relative thickness of the ejecta, the profiles display the full range of those observed.  Composite contributions, e.g. thin wall bipolar ejecta, are indicated by the narrow emission spikes in the forbidden profiles.  Similar profiles were reported in Iijima (2012) for the recurrent nova CI Aql.\footnote{A more complete discussion found in Shore et al. (2012, {\it submitted to A\&A}, dealing with T Pyx 2011}.  
\begin{figure}
\centerline{\includegraphics[width=11cm]{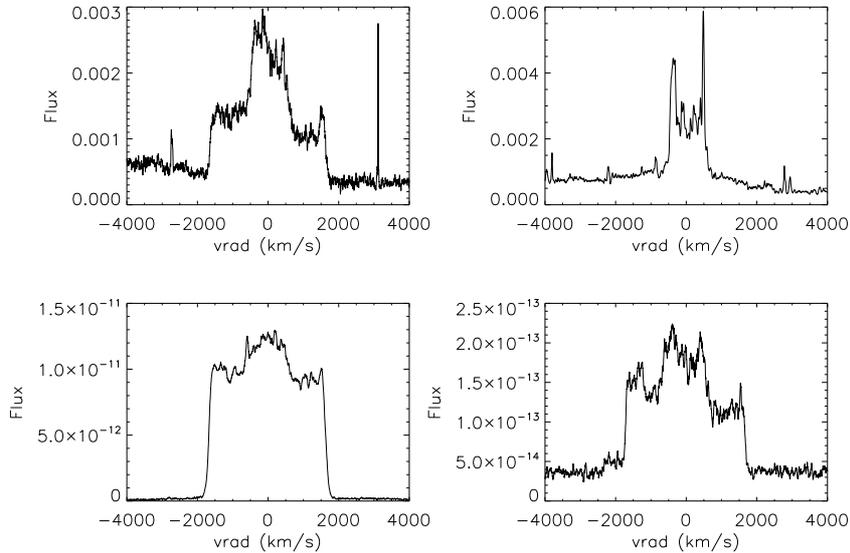}}
\caption{T Pyx 2011: Variations of [N II] 5577\AA\ (left) and N IV] 1486\AA\  (right) during the optically thin stages of the outburst (2011 Oct., top,  and 2012 Apr., bottom).\label{f:one}}
\end{figure}

As mentioned in sec. 3.2, the narrow absorption lines observed during the Fe-curtain phase also show the presence of structure long before the emission lines appear.  These are organized within the larger scale structures inferred from both the later nebular profiles and also by weak emission features detectable on the transition stage line profiles.  There are a few cases, notably (again) T Pyx 2011, where the broad absorption troughs have mirror correspondences in the emission part of the line profiles.  These become more evident as the optical depth decreases (see Fig. 4), especially on the hydrogen line profiles.   What structures the ejecta, why they have this predominantly bipolar symmetry, is a question posed by the interferometric observations (see Chesneau \& Banerjee, this issue).  For systems like RS Oph and V407 Cyg, this is easy to understand, the explosions take place off-center in a dense wind of a companion giant that has a strong density gradient (at least inverse square).  In addition, orbital angular momentum insures that the giant's wind is concentrated toward the orbital plane.  The mass gaining WD is likely also surrounded by a disk but probably this is not in itself is not significant.  Instead, the spin-up of the white dwarf by accretion, the rotational angular momentum, must be playing a role.  This would explain why the symmetry is seen also in``normal'' systems, i.e. those with compact companions.    But the mass of the disk is not be completely negligible for recurrent novae.

\begin{figure}
\centerline{\includegraphics[width=11cm]{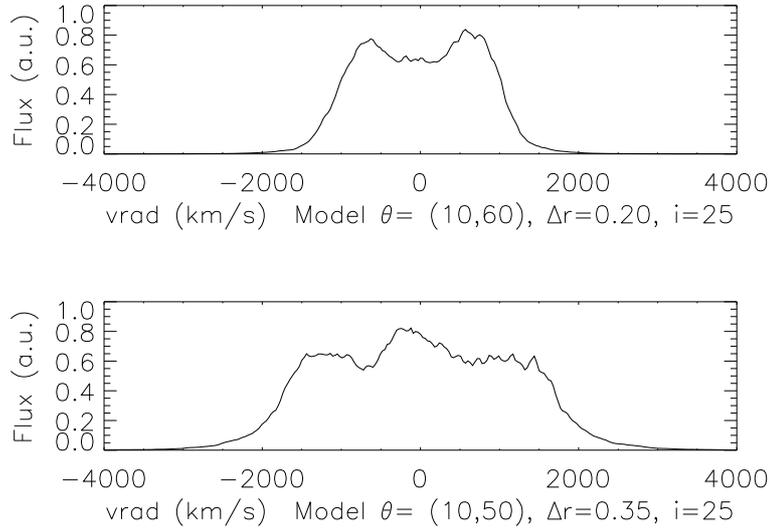}}
\caption{Expected line profiles for different bipolar ejections.  The two angles are the inner and outer polar angles and $\Delta R$ is the relative thickness of the shell (the minimum fractional velocity relative to the maximum ejecta radial velocity).  The angle of inclination, $i$, was chosen to match the optical interferometry and orbital characteristics of this particular system.\label{f:one}}
\end{figure}

\begin{figure}
\centerline{\includegraphics[width=14cm]{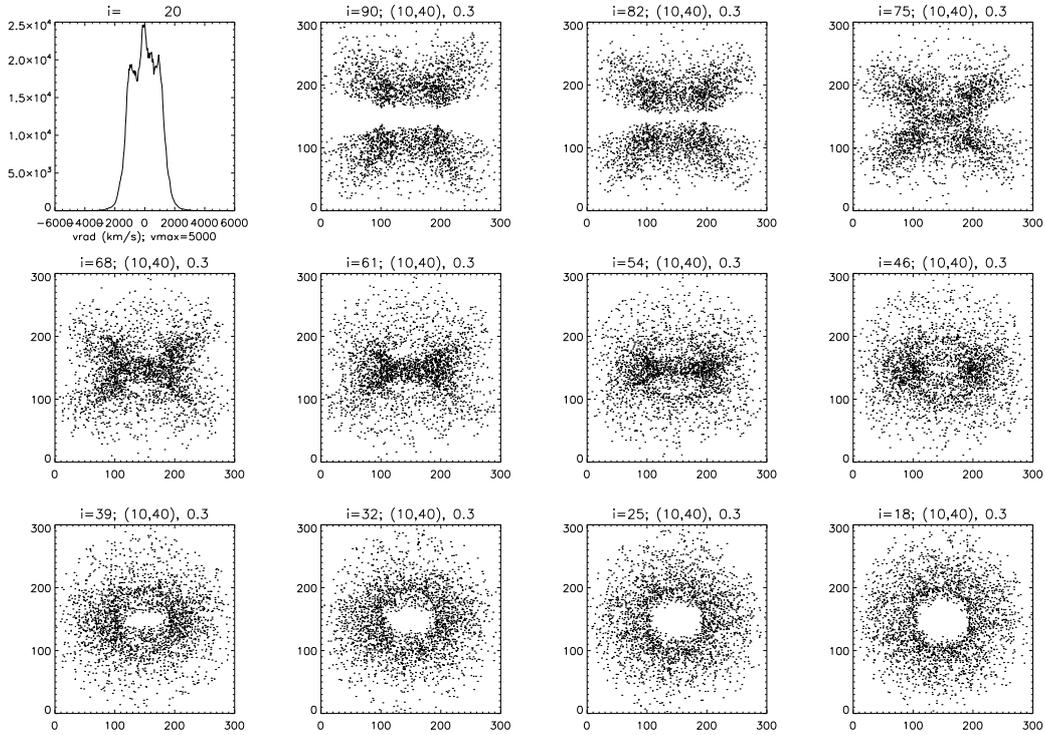}}
\caption{Model bipolar structures with a sample line profiles, computed for the T Pyx 2011 outburst.  Compare with Fig. 10.  The images can be compared with the HR Del optical image in Fig. 13.  The range of images is for different inclination angles, all other parameters were kept fixed.\label{f:one}}
\end{figure}

\begin{figure}
\centerline{\includegraphics[width=7cm]{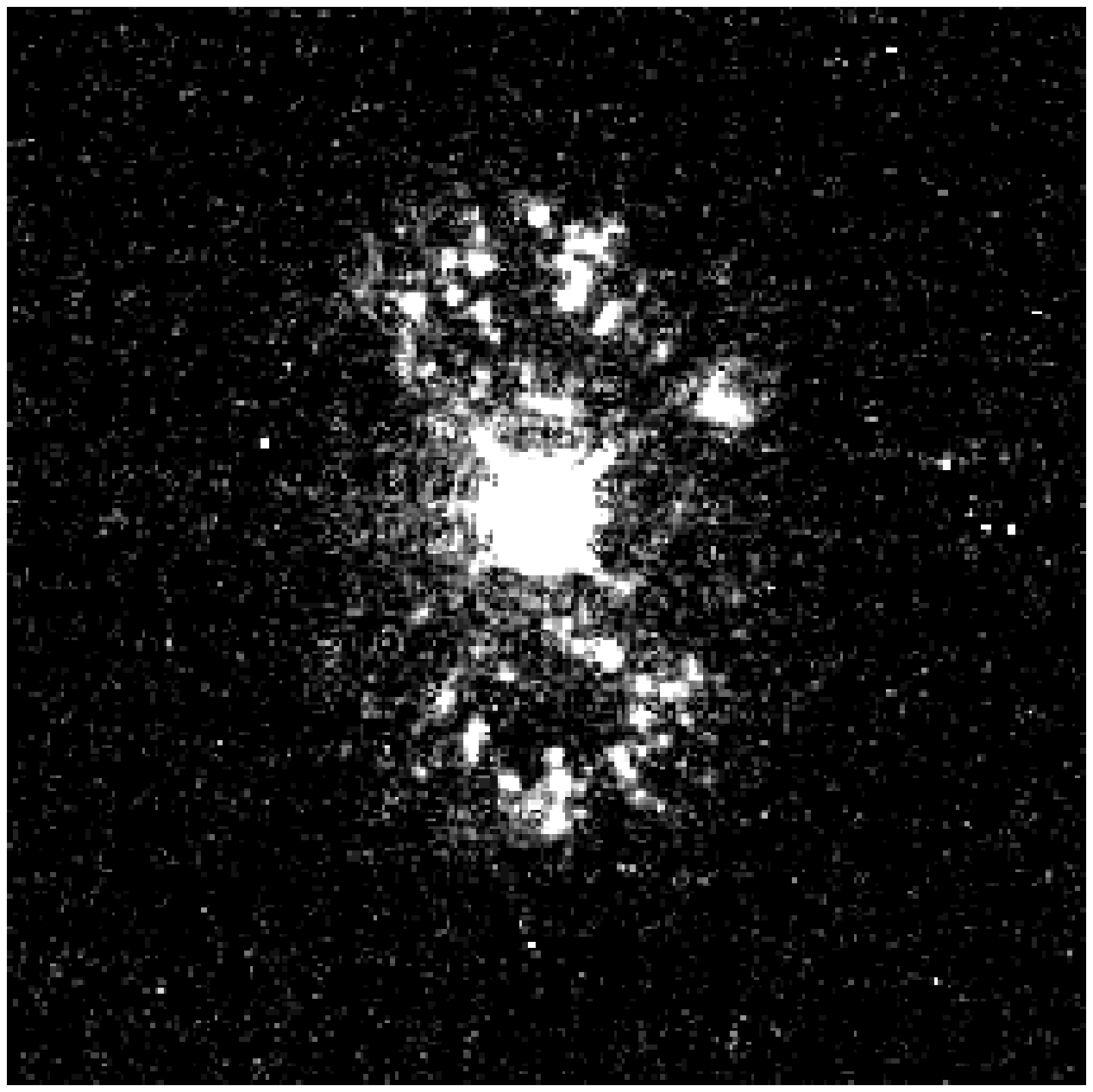} \qquad
            \includegraphics[width=7cm]{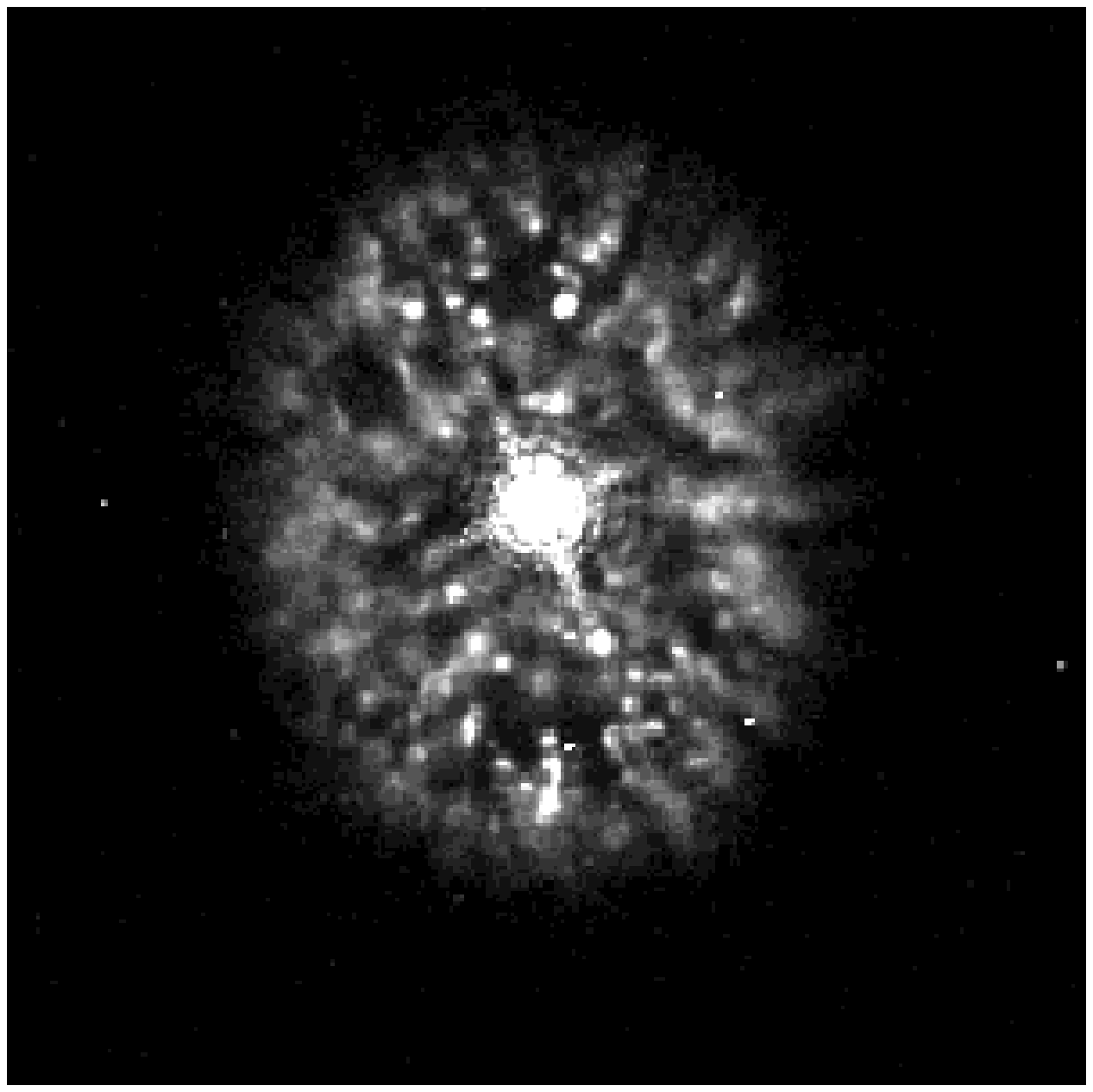}}
\medskip
\centerline{\includegraphics[angle=0,width=7cm]{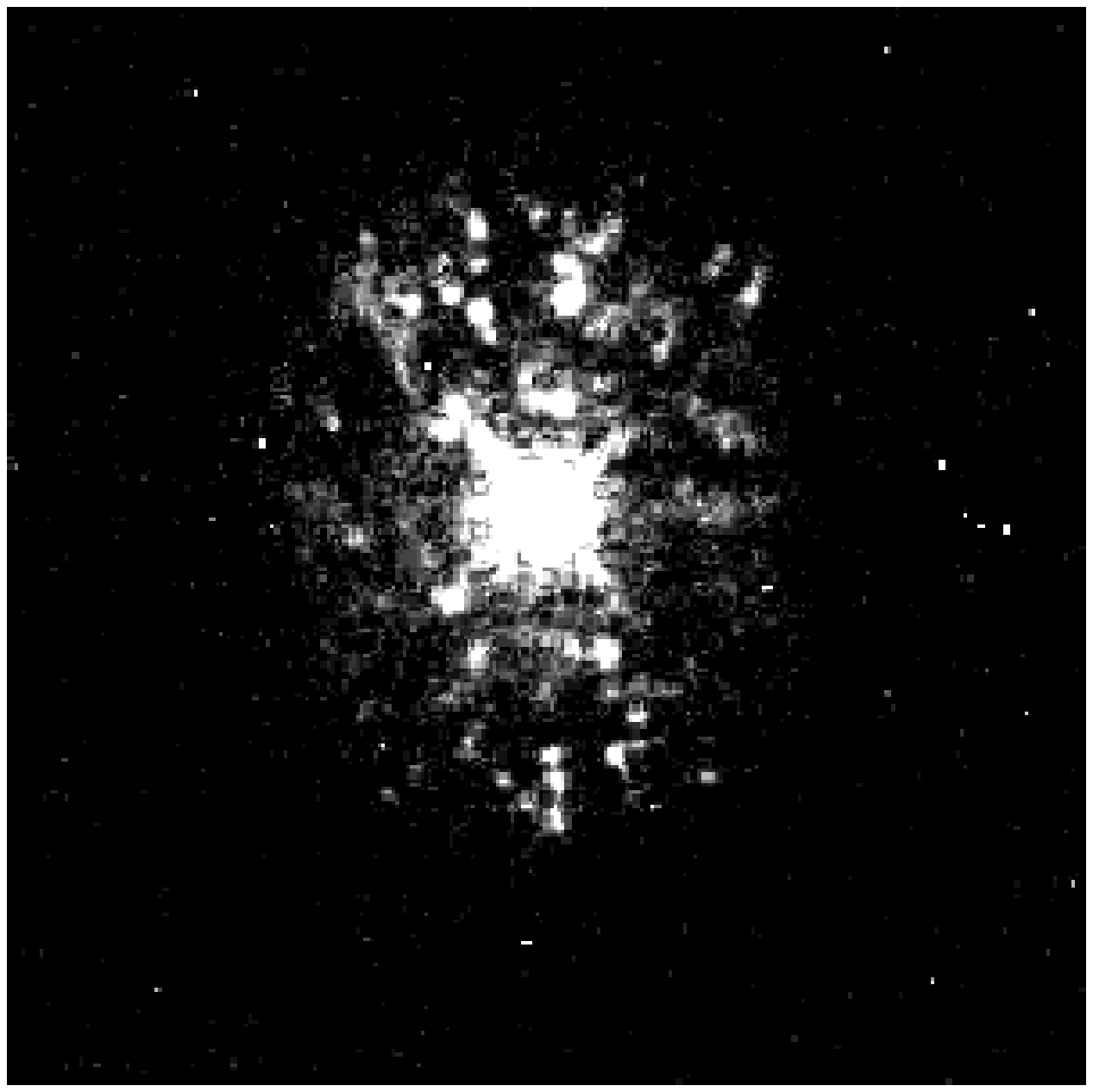} \qquad
            \includegraphics[origin=ltangle=0,width=5cm]{hrdel-nii.ps}}
\caption{The ejecta of HR Del 1967 imaged with WFPC2/{\it HST} in 1997 and 1998.  Top left: [O III] 4959+5007\AA; top right: H$\alpha$, bottom center: [N II]6583\AA.  The differences in the imaging match the expectations of the ejecta structure derived from modeling the line profiles (see Harman \& O'Brien (2003) for further examples of the modeling and a discussion of the previous literature on this nova). \label{f:many}}
\end{figure}
\section{Winds?}

One basic question remains unanswered at this time: what about winds?  Is there evidence for some sort of continuous mass loss during or after the explosion?  This problem has exercised the community for decades.  The issue is complicated by the fact that, when optically thick, it is very difficult observationally to distinguish a photosphere in a freely expanding medium and a stationary surface in space through which material flows based only on the light curve.  Since that is what most discussions have been based on, the wind models have relied on {\it ad hoc} mechanisms for inducing the time variations.  For instance, as is well known from Luminous Blue Variables (e.g. HDE 268858 in the LMC, AG Car in the Galaxy), when the mass loss increases such that the column density exceeds about $10^{23}$cm$^{-2}$, the wind recombines and a spectral redistribution of flux produces a rise and extended period of maximum light.  Both the color and, it appears, excitation temperatures drop below about 15 kK, as also seen during the optical peak of classical novae.   But there are two important distinctions.  The absorption lines are generally not saturated at the terminal velocity, or if they are they rapidly change to become a continuous absorption profile.  The second is the development of those profiles that weaken systematically in time.  It is certainly plausible that a wind is present post-ejection, likely during the supersoft phase when the XR luminosity is near its maximum and radiation pressure suffices to drive an outflow.  But this is {\it not} optically thick and does not seem to be involved with the same manifestations we observe during the early stages of the nova outburst.

\section{Phenomenological classifications}

\subsection{The CTIO classes}

The previous description of the physical processes can be codified by a simple scheme worked out by Williams and collaborators (Williams 1992a,b; Williams et al. 1994).  There are three classes of optical nova spectra in this taxonomy: the Fe, HeN, and hybrid groups.  The first is low ionization, with both emission and absorption by the iron peak elements (mainly Fe) in forbidden and permitted lines, with or without P Cyg profiles.  The second lacks those lines but, instead, is dominated by emission lines of light elements, mainly helium and nitrogen as the name implies.  The third is the ``changeling'' group that start as Fe type and end as HeN novae.   

In light of the foregoing discussion, you see that these are variants on a theme: the stages of changing transparency of the ejecta correspond to these types {\it in succession}.  That some novae are caught late, in the sense that they have already become optically thin, or stay opaque for a long time, is in a sense irrelevant.  Some very low ejecta mass systems, especially recurrent novae (see below), pass through the opaque stages so quickly that the are frequently first observed as HeN types.  Others get ``stuck'' in the Fe stage and, when that clears, pass directly to the coronal and nebular stage.  But it is important to emphasize that there is nothing distinctive about these types, they are indicative of the expansion stage of the nova and nothing else.  

\subsection{A remark on the photometric taxa and phenomenology}

The same may be said of the timings used for the optical light curves.  Introduced by Payne-Gaposchkin (1957) as a convenient way of separating the various taxa of photometric behavior, the two parameters $t_2$ and $t_3$ -- the time to fall two or three magnitudes from peak -- has taken on a great significance even if there is little that is particularly physical in the distinction.  So-called ``fast'' novae have $t_3$ less than about 10 days, obviously slow novae are the longer group.  An illustration of the imprecision of this interpretation is that several symbiotic novae and $\eta$ Car were considered ``slow novae'' in this sense.  We know now they're neither slow nor novae.  Seven, more or less, distinct light curve morphologies have been distinguished by Strope et al. (2010) but without linking the spectroscopic stages to the particular photometric peculiarities.  There are certainly important distinctions, e.g. oscillations in the decline phase, long duration maxima (months), dust forming events, and ensembles of peaks during maximum light (jitter), but how these relate to structural features of the ejecta or the behavior of the central source remain unsolved problems.

One of the most frequently used photometric tools, the Maximum Magnitude Rate of Decline (MMRD) relation, finds its explanation in the flux redistribution during the curtain phase.  Again, this is not a sure indicator of either a wind or freely expanding ejecta.  It simply requires that during the UV opaque stage the central source remains approximately constant .  This is the maximum magnitude part of the relation.  The rate of decrease in the optical depth of the ejecta depends on the mass and maximum expansion velocity of the gas.  Thus, a correlation between the ejecta mass, maximum velocity (in other words, the kinetic energy), and luminosity --  for {\it any} such event -- insures the existence of some kind of MMRD relation.\footnote{For example, it would be interesting to see if there is a similar MMRD relation for Luminous Blue Variables, for which the optical event is connected with a sudden increase in the mass outflow through a strong stellar wind.  These outbursts are  analogs of novae in slow motion, at least for the details of the spectral formation and even the stages through which they pass.  The velocities are usually lower an they are winds rather than explosions, but there may be some similarity.  On the other hand, being winds, the differences they show in outburst compared to novae are relevant to the question of whether or when winds form in classical novae.}   There is , however, another rather nasty issue let hanging.  This all assumes that the ejecta are completely covering the central source, and that the filling factor of any flux redistributing matter is very high.   In this picture, it is only necessary that the UV opaque parts of the ejecta re more nearly spherical to create the bolometric effects; this is a matter for further investigation.

\subsection{The distinction between ``classical'' and recurrent novae}

Here  human bias enters.  We don't have the longevity, neither biological nor archival, to know whether a system has repeatedly erupted with a cadence of longer than once every few centuries; see Anupama (2008).   As Schaefer has repeatedly stressed (e.g. Schaefer 2010), there are also instances of likely missed outbursts for some known repeat offenders, when the peak was missed because of meteorological or astronomical constraints.  Two recent examples suffice to show the truth of this assertion. KT Eri was discovered on the decline but its peak was ``recovered'' using satellite photometry from {\it SMEI} satellite, since it was in the field of view of the imager although close to the solar disk (Hounsell et al. 2010).  The spectacular example of N Mon 2012, first discovered as a {\it Fermi}/LAT transient in early June 2012, was only found in outburst observationally two months later when the region was again visible from the ground, because of solar constraints, at which time the spectra showed it to be already in the nebular transition stage.  Recurrents have generally such rapid declines, because of their low ejecta masses and high expansion velocities, that the likelihood of missing any single outburst is not negligible.   On the other hand, the interval between eruptions is system-specific and not predictable, as the 2010 nova eruption of the symbiotic Mira V407 Cyg demonstrated.\footnote{It is important to note that this system had no previously recorded {\it nova} event.\footnote{Even more dramatic is the example of T Pyx whose 2011 outburst was awaited after it last exploded in 1966.  Generating an enormous literature, it was one full cycle late in eruption.}  There was a symbiotic nova-like event in the 1930s but that was completely different.  The 2010 event was a true explosion, indistinguishable in its various stages from RS Oph 1985 or 2006.  The identification with a recurrent nova of the symbiotic-like variety is a stretch but justified  by one hard {\it fact}: no classical nova, whose ejecta mass is the same as the non-recurrent systems, has ever been observed among the symbiotic-like recurrent novae, see below.}  This is an area that has too many open issues, the most central of which is what happens in the long interval between centuries and tens of thousands of years.  Thus,  it seems premature to rigorously separate novae into different types based only on the repetition rate.  
 
\subsection{The physically distinct subgroups: CO and ONe subtypes}

The passage to transparency renders the analysis of abundances, structure, and physical properties simpler for novae since the whole ejecta are visible.  Two, apparently distinct, white dwarf progenitors have been distinguished based on the abundances as preserved in the ejecta.   The majority of explosions we see among classical novae happen on CO white dwarfs.  A separate group, less numerous although perhaps over-represented in the discovery statistics because of their energetics, are from ONe WD progenitors.  These {\it may} both produce recurrent nova outbursts, that is still an open question of great importance, although at this point the weight of evidence is toward the CO progenitors (for more discussion see the review by Starrfield et al. in this issue).    

But there is a much more important ``fact'': the ONe novae seem to be remarkably spectroscopically homogeneous.  So close are they (see, e.g. Fig. 8) that they can be used as templates for planning observations.   A more significant feature is that, like SN Ia, they may be a sort of standard candle (or nearly so).  There are two features in which these novae are strikingly similar and unique.  It is not just the presence of the [Ne IV] 1602\AA\ and [Ne V] 1575\AA\ UV lines that sets them apart.  They pass through a very particular transitional stage after the Fe-curtain clears: a genuine continuum (not merely myriad blended lines)  plus P Cyg profile stage that resembles an O star spectrum.  This is very shortlived but has been observed well in LMC 1990 Nr. 1, V1974 Cyg (Shore et al. 1993; Moro-Mart\'in et al. 2001), V382 Vel, and LMC 2000 (Della Valle et al. 2002; Shore et al. 2003).  Others, with more fragmentary coverage, also seem to show it.  It is most evident in the UV on the strong resonance lines of the usual suspects, C IV 1550\AA\ and Si IV 1400\AA\ but it also corresponds to an optical continuum.  The second is the structure on the emission lines, an example is shown in Fig. 8 but also in Fig. 16 for a {\it suspected} ONe nova Mon 2012, follows a sequence.  The [O I] 6300\AA\ line, for instance, was very different in V382 Vel and, say, OS And 1986 (a CO nova), in showing a complex filamented profile (e.g. Della Valle et al. 2002).   

\section{The extreme case of the symbiotic-like recurrent novae (SyRNe)}

Among the recurrent novae, there is a distinct subgroup of WDs whose companions are evolved giants instead of compact main sequence (or nearly) stars.  This group, T CrB, RS Oph, V3890 Sgr, V745 Sco, and the recently proposed V407 Cyg, are -- above all -- distinct for their environments (see Evans et al. (2008), on the RS Oph 2006 event, and references therein) .  Any explosion in such a medium will evolve essentially differently than freely expanding ejecta.  First, there is another source to power the emission: the shock formed by the ejecta plowing through the dense wind reaches temperatures well above 1 MK for the characteristic velocities of the WD.  Second, as the shock expands it accelerates and, consequently, engulfs wind material that has a different composition.  Within about one week, the ejecta will normally have accumulated roughly their own mass in the traverse, hence the initial composition has been altered in ways that depend on the abundances of the giant.  Third, the environment is neither uniform nor spherical, hence neither is the shock.  Its velocity is no longer simply ballistic, and the line profiles will depend on the the specific density and temperature of the locale and the radial velocity with respect to the observer.  Even for the Sedov phase\footnote{This is the early stage of the blast wave  when the ejecta go from free expansion to self-similar expansion and adiabatic acceleration with constant energy $E_0$ of the ambient medium of density $\rho_0$ such that the outer radius varies as a power law in time, $R(t) \sim t^n$ where $n$ depends on the radial variation of $\rho_0$ (see Bode \& Kahn 1985; Sokoloski et al. 2006; Walder et al. 2008; Orlando et al. 2009; Shore et al. 2011a)} there is a much more rapid expansion of exiting material than the side of the shock directed inward toward the companion.  Last, the ionization of the wind is governed by a combination of the shock emission and that of the WD and this causes the profiles to vary in a much more complicated way than the freely expanding case.  An example of the complexity of the line profiles, and their disentanglement, is shown for V407 Cyg in Fig. 14.

\begin{figure}
\centerline{\includegraphics[width=11cm]{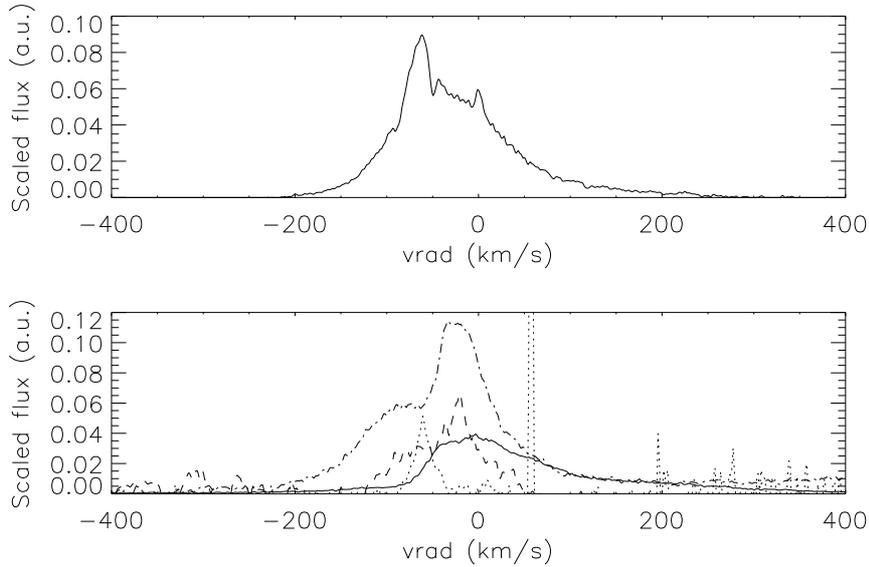}}
\caption{A late-time optical spectrum of V407 Cyg illustrating how to empirically  separate the contributions of different physical processes and spatially distinct components to the observed line profile (2011 Aug. 21 NOT spectrum).
Top: [O I] 6300\AA. Bottom: [O III] 5007\AA\ (solid, ambient ionized wind), Mg I 4571\AA\ 
(dot; chromosphere and inner neutral wind), [S II] 6716\AA\  (dash; shocked gas), [N II] 6548\AA\ (dot-dash; extended lower ionization outer wind), from Shore et al. (2012).\label{f:one}}
\end{figure}

Line profile changes are also attributable to a time dependent ionization front that propagates outward in the ambient wind.    One of the components shown in Fig. 14 is due to this.  The Balmer lines are especially good optical probes of the phenomenon.  The low velocity absorption component, produced (as in symbiotic stars) from the line of sight to the WD seen through the wind, decreases in intensity as the matter ionizes.  For instance, as shown in Fig. 15, the absorption on the higher series members can actually disappear, being replaced by emission from recombination at the same velocity.  The velocity of the strongest absorption also changes systematically as the absorption becomes weighted toward the chromospheric contributed by the giant.  On the resonance lines in the UV, the absorption may completely vanish (this was noted, for instance, during the 1985 outburst of RS Oph, see e.g. Shore et al. (1996) for the Fe-peak absorption curtain against the C IV lines from the ejecta.  No such data were available for the 2006 outburst).  After shock breakout, when the ejecta effectively emerge from the wind, parts of the wind recombine.  This was noted in the 2010 V407 Cyg event, especially for the [N II] 5755, 6548, 6583\AA\ lines.  It is important to never forget that the spectroscopic evolution of the SyRNe is completely dominated by this complex radiative and mechanical interaction between the hot post-explosion WD, the expanding shock, and the wind of the companion and is {\it distinct} from that expected for a classical nova.

\begin{figure}
\centerline{\includegraphics[width=11cm]{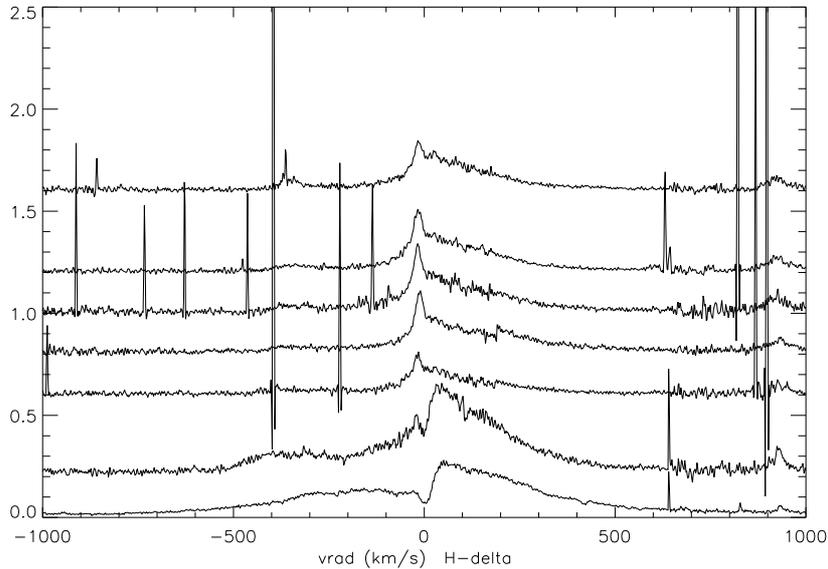}}
\caption{Variations in V407 Cyg of the H$\delta$ line during outburst. Time proceeds from the bottom up (from Shore et al. 2011a).  In this case, as the ionization front propagated ahead of the expanding shock, within the wind, previous absorption components on this Balmer line went into emission.  At the same time, which the wind absorption weakened on H$\alpha$ and H$\beta$ it did not disappear during more than one year after the outburst.\label{f:one}}
\end{figure}

Symbiotic stars also display a particular phenomenon not seen in most other stellar atmospheres.  Two strong, broad emission lines, at 6825\AA\ and 7082\AA\ that were first described in the 1940's were finally explained by Schmid (1989) to be the Raman scattering of O VI 1036,1042\AA\ by Ly$\beta$ 1025\AA.  The two emission lines are formed in the neutral part of the red giant wind in these systems, the high ionization line arising from the region around a hot WD companion.  The 6825\AA\ line also observed in RS Oph 1933  by Joy \& Swings (1945) who, however, preferred the identification of [Kr III].   Recently its Raman scattering feature identification has been confirmed by Iijima (2009) during the 2006 outburst.   This is a phenomenon, along with Rayleigh scattering from the neutral wind, that does not occur in other classes of novae and not all SyRNe show it either (e.g. it was not observed during the 2010 outburst of V407 Cyg, Shore et al. 2011a).  These lines are particularly important since they are intrinsically optically thin and, being formed by scattering, are sensitive probes of the geometry of the binary (for instance, relative orbital phase of the WD and giant) and the structure of the red giant wind.

In summary, the spectroscopic differences between these novae and any others are caused by the environment, not by the explosion itself.  Whatever part of the wind is not overtaken by the shock is irradiated by both the shock-produced XRs and the WD.  The resulting ionization front, that propagates through the red giant wind, produces a rich emission line spectrum including fluorescent transitions that are pumped from the UV resonance states.  These three components, the chromosphere, wind, and shock, are easily distinguished by their profiles.  The wind lines dominate the late spectra of these objects but are the narrow components present throughout the visible region.  Those from the shock depend on the relative orientation of the front to line of sight through the wind.  A portion of the shock, indeed also of the ionized wind,  is  blocked from view by the red giant and the inner wind.  The lines are asymmetric, which is mainly from the difference in the velocity of the inward and outward (or off planar) parts of the front.  There may also be a contribution from the giant's chromosphere that is shielded by the star from the ambient radiation.  These are shown in the sample spectra from the V407 Cyg  2010 outburst.   These features are so distinct that they can even be separated at low resolution (e.g.  Gonzales-Riestra (1992) for V3890 Sgr and discussion in Shore 2008).   One final point.  The abundances obtained from these ejecta are also affected by the passage through the companion's wind.  As discussed by Walder et al. (2008), within about a week the ejecta have roughly doubled in mass (or more, depending on the  companion and the orbital characteristics).  What is seen in the later stages, when the spectra become nebular, will have been contaminated with matter from the red giant.  This may explain why, for instance, for the compact recurrent systems, like U Sco and LMC 1990 Nr. 2, the He/H ratio is larger than solar yet is closet to normal for the SyRNe (e.g. Anupama \& Prabhu 1989).  As a last point, unlike classical novae, the line of sight through the red giant changes significantly the appearance of the emission lines and may even obscure a part of the event (both because of occultation by the companion itself and also absorption through the inner wind, this also affects the interpretation of the post-nova system).

\section{Abundance determinations and pitfalls}

Novae shells are not static H II regions and there are phases at which they are far from the usual photoionization equilibrium conditions usually found in the interstellar medium.  To date, all studies have used integrated line fluxes to obtain the chemical composition of the ejecta but there are many indications that this may not be the best approach.  For example,  UV spectra of V1974 Cyg indicated variable Ne/C and Ne/He ratios in the knots of the spatially resolved ejecta (Shore et al. 1997).  The dispersions in derived abundances obtained from meta-analyses of mutiwavelength spectra (e.g. Schwarz et al. 
2007, Vanlandingham et al. 2005)  are frequently larger than the errors in the measured fluxes.  These variances have long been simply quoted as statistical uncertainties but they may be indicating something very important, that the elements are not homogeneously distributed throughout the ejecta.  In addition to the large variations in the ejecta geometries displayed by the different characteristic shapes of the profiles, the flux differences in the same velocity interval of the profile of different species {\it may} be signaling real abundance differences.  To study this requires high signal to noise data at high (at least 10 km s$^{-1}$ velocity resolution) since the emission knots usually have widths of $<$100 km s$^{-1}$ with fair temporal  cadence so the effects of the changing physical conditions can be properly separated from the chemical differences.     The range of optical depths present at any time in the expansion is well illustrated by the variation in the Balmer line profiles in the early stages of the event (see Fig 16 for Mon 2012 at about 60 days after maximum V).  The changes in the profiles also show how the geometry inferred from one set of transitions may be misleading even fairly late in the expansion (in this case during the transition stage).

\begin{figure}
\centerline{\includegraphics[width=12cm]{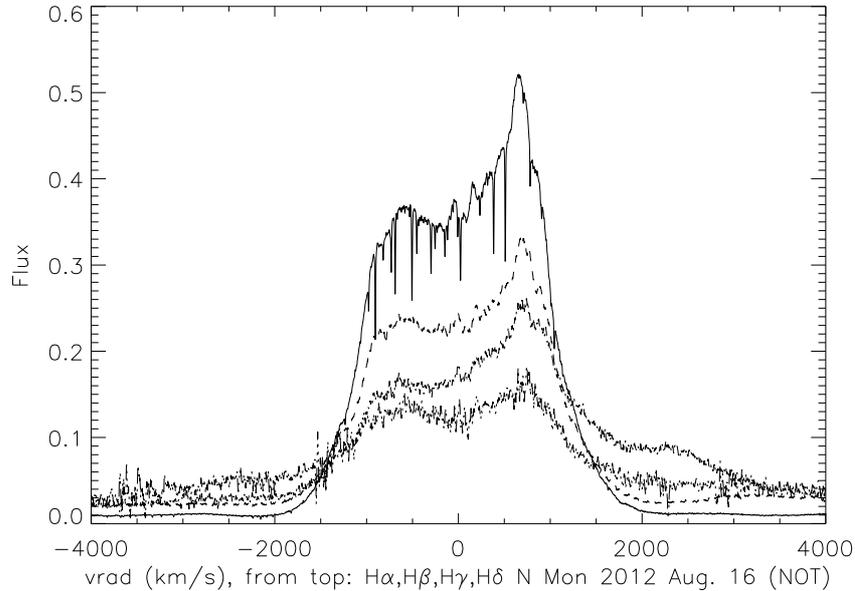}}
\caption{The Balmer lines (H$\alpha$-H$\delta$) in nova Mon 2012 about 3 months after probable optical maximum (this nova was first discovered after a period of Sun constraints).\label{f:one}}
\end{figure}

Several other problems affect analysis of novae that do not enter other environments with apparently similar spectra.  Being time dependent, the expansion time may dominate the recombination rate, throwing the system out of photoionization equilibrium.  Although useful, indeed standard, photoionization codes must be applied to individual spectra with the constraint that they produce the same abundance set.   A second important issue is the effect of large density contrasts within the ejecta.   This is not the same issue as large scale density gradients.  Those can be dealt with straightforwardly by a variable optical depth and escape probability.  Instead, the meaning of ``filling factor'', an almost magical parameter that converts the observed luminosity into an emitting mass, is physically subtle.  The usual representation is of a medium only partly filled with matter, some fraction being empty and, consequently, transparent and not contributing to the observed emission.  This is not the same thing one sees in nova ejecta where there are large density contrasts even in small velocity ranges within the profiles.  These knots, having different densities, also radiate differently and, more importantly, recombine differently.  Integrated line emissivities cannot capture this and the mean abundances derived from individual lines may be misleading.  Even in ensembles, since the different ionization states are weighted to different parts of the ejecta, there may be a large spread in the resulting abundances for any element.  This has always been considered as an error but, as with the issue of chemical homogeneity, the nonlinear dependence of the emissivity on density and the differences in velocity of individual knots may produce large dispersions in the derived abundances.  To date there have been very few attempts to combine forward ejecta modeling with Monte Carlo photoionization codes (e.g. Ercolano et al. 2003) to study such effects; this remains work for the future.

One way of thinking about this is the sensitivity of recombination to density, $n_e$.  The emissivity varies as $n_e^2$, hence small fluctuations will amplify the effect, while the recombination time is linear.   Were there only a smooth linear velocity gradient, the characteristic recombination time for each density that scales as $t_* \sim n_e^{-1/3}$.  But if at the same velocity different knots have different densities, even for a constant mass shell, there will be a range of recombination timescales and ionizations present.  The expansion time is independent of position in the ejecta in the ballistic case so, in principle, there should be a single time at which the ejecta recombine.  But if there is a dispersion of densities at a given velocity and those advect self-similarly, the recombination will take place in each at a slightly different rate.  There may be no need to hypothesize shielding of parts of the ejecta from FUV radiation to produce significant ranges in ionization (including neutrals) even in late-time spectra.

\begin{figure}
\centerline{\includegraphics[width=12cm]{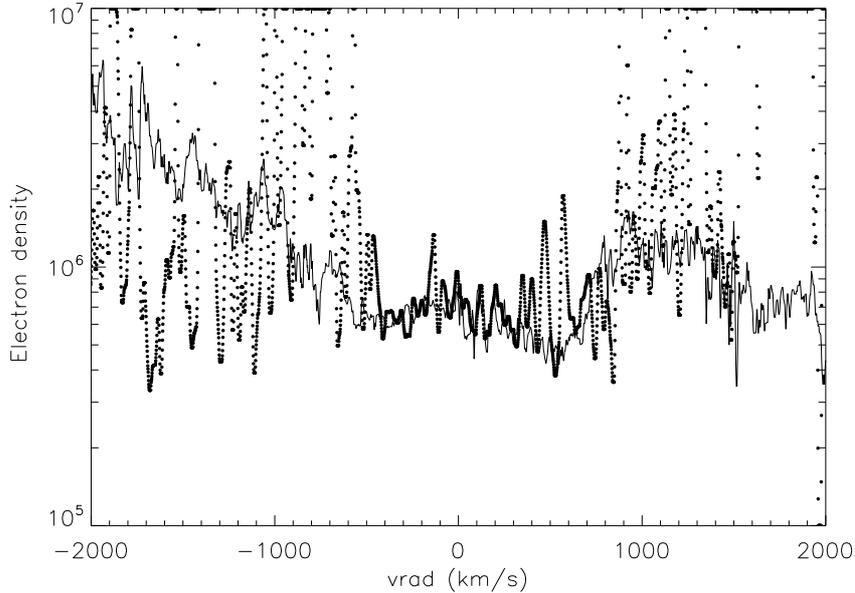}}
\caption{The [N II] (solid) and [O III] (dot) line diagnostics of the electron density for T Pyx from 2012 Apr. (from NOT spectra, R$\approx$ 64000) the late nebular phase.  The electron temperature was determined independently from knowing that the line profiles of the two ions are identical.\label{f:one}}
\end{figure}

To overcome some of these difficulties requires both high signal to noise and high spectral resolution.  The lines, once optically thin, can be analyzed in a manner analogous to that use for the interstellar medium (see, e.g. Savage et al. 1991).  An example is shown in Fig. 17  for the recurrent nova T Pyx during its latest (2011) outburst.  The structures on the line profiles in different spectral regimes persist through the expansion (e.g. UV vs. infrared in V1974 Cyg, Hayward et al. 1996).  Since the different parts of the spectrum become transparent at different times, beginning in the near infrared,  the ejecta can be probed at similar geometric depths at different times.  This not only checks for the durability of the emission knots but also allows an assessment of the physical conditions through sometimes subtle changes in the profiles.   One case, the symbiotic-like system V407 Cyg 2010, shows this to particular advantage for the recombination of the envelope of red giant after the shock had passed breakout (e.g. Shore et al. 2011a, 2012).  Differential changes in line profiles have been noted sporadically in the literature but have not been fully exploited.  It is now feasible to extend this method by matching many transitions of different species in multiwavelength studies.   

\section{Final remarks}

After more than a century, many of the puzzles have finally been resolved, that we have finally passed out of the phenomenological era to face the most basic questions of these remarkable cosmic objects.  The vista is far richer and more broadly applicable than any of the ``founders'' could have dreamed.

\section*{Acknowledgements}

This review is written near the end of an epoch of discovery that started twenty years ago with the outburst of nova V1974 Cyg, the first truly multiwavelength, high resolution campaign.  Now, so many years later, as we face the end of that era with the approaching closing of the ultraviolet spectroscopic window.  But with the new discovery of very high energy $\gamma$-ray emission from classical novae, we're starting anew.  For discussions, collaborations, and critiques and fisticuffs over the  years, I warmly thank G. C. Anupama, Jason Aufdenberg, Thomas  Augusteijn, Mike Bode,  Jordi Casanova, Teddy Cheung, Nye Evans, Bob  Gehrz, Peter Hauschildt, Jordi Jos\'e, Pavel Koubsky, Jan-Uwe Ness, Julian Osborne, Kim Page, Greg Schwarz, George Sonneborn, Sumner Starrfield, Karen Vanlandingham, and Glenn Wahlgren.  I also thank the recent recruits to this business with whom I am proud to have worked during their studies at Pisa: Diana Di Nino, Ivan De Gennaro Aquino, Katia Genovali, and Walter del Pozzo.  Some of the studies described here have been, in part, supported by NASA, OPTICON, and the INFN.  Finally, I thank the editors of this special issue for their kind invitation and patience with the delays in the delivered manuscript. 


\appendix
\end{document}